\documentclass[letterpaper,compsoc,twoside]{IEEEtran}
\usepackage{fixltx2e} % LaTeX patches, \textsubscript
\usepackage{cmap} % fix search and cut-and-paste in Acrobat
\usepackage{ifthen}
\usepackage[T1]{fontenc}
\usepackage[utf8]{inputenc}
\usepackage{amsmath}

\usepackage[font={small,it},labelfont=bf]{caption}
\usepackage{float}

\usepackage{longtable,ltcaption,array}
\newlength{\DUtablewidth} % internal use in tables

\pdfoutput=1
\usepackage{scipy}
\makeatletter
\def\PY@reset{\let\PY@it=\relax \let\PY@bf=\relax%
    \let\PY@ul=\relax \let\PY@tc=\relax%
    \let\PY@bc=\relax \let\PY@ff=\relax}
\def\PY@tok#1{\csname PY@tok@#1\endcsname}
\def\PY@toks#1+{\ifx\relax#1\empty\else%
    \PY@tok{#1}\expandafter\PY@toks\fi}
\def\PY@do#1{\PY@bc{\PY@tc{\PY@ul{%
    \PY@it{\PY@bf{\PY@ff{#1}}}}}}}
\def\PY#1#2{\PY@reset\PY@toks#1+\relax+\PY@do{#2}}

\expandafter\def\csname PY@tok@gd\endcsname{\def\PY@tc##1{\textcolor[rgb]{0.63,0.00,0.00}{##1}}}
\expandafter\def\csname PY@tok@gu\endcsname{\let\PY@bf=\textbf\def\PY@tc##1{\textcolor[rgb]{0.50,0.00,0.50}{##1}}}
\expandafter\def\csname PY@tok@gt\endcsname{\def\PY@tc##1{\textcolor[rgb]{0.00,0.27,0.87}{##1}}}
\expandafter\def\csname PY@tok@gs\endcsname{\let\PY@bf=\textbf}
\expandafter\def\csname PY@tok@gr\endcsname{\def\PY@tc##1{\textcolor[rgb]{1.00,0.00,0.00}{##1}}}
\expandafter\def\csname PY@tok@cm\endcsname{\let\PY@it=\textit\def\PY@tc##1{\textcolor[rgb]{0.25,0.50,0.56}{##1}}}
\expandafter\def\csname PY@tok@vg\endcsname{\def\PY@tc##1{\textcolor[rgb]{0.73,0.38,0.84}{##1}}}
\expandafter\def\csname PY@tok@m\endcsname{\def\PY@tc##1{\textcolor[rgb]{0.13,0.50,0.31}{##1}}}
\expandafter\def\csname PY@tok@mh\endcsname{\def\PY@tc##1{\textcolor[rgb]{0.13,0.50,0.31}{##1}}}
\expandafter\def\csname PY@tok@cs\endcsname{\def\PY@tc##1{\textcolor[rgb]{0.25,0.50,0.56}{##1}}\def\PY@bc##1{\setlength{\fboxsep}{0pt}\colorbox[rgb]{1.00,0.94,0.94}{\strut ##1}}}
\expandafter\def\csname PY@tok@ge\endcsname{\let\PY@it=\textit}
\expandafter\def\csname PY@tok@vc\endcsname{\def\PY@tc##1{\textcolor[rgb]{0.73,0.38,0.84}{##1}}}
\expandafter\def\csname PY@tok@il\endcsname{\def\PY@tc##1{\textcolor[rgb]{0.13,0.50,0.31}{##1}}}
\expandafter\def\csname PY@tok@go\endcsname{\def\PY@tc##1{\textcolor[rgb]{0.20,0.20,0.20}{##1}}}
\expandafter\def\csname PY@tok@cp\endcsname{\def\PY@tc##1{\textcolor[rgb]{0.00,0.44,0.13}{##1}}}
\expandafter\def\csname PY@tok@gi\endcsname{\def\PY@tc##1{\textcolor[rgb]{0.00,0.63,0.00}{##1}}}
\expandafter\def\csname PY@tok@gh\endcsname{\let\PY@bf=\textbf\def\PY@tc##1{\textcolor[rgb]{0.00,0.00,0.50}{##1}}}
\expandafter\def\csname PY@tok@ni\endcsname{\let\PY@bf=\textbf\def\PY@tc##1{\textcolor[rgb]{0.84,0.33,0.22}{##1}}}
\expandafter\def\csname PY@tok@nl\endcsname{\let\PY@bf=\textbf\def\PY@tc##1{\textcolor[rgb]{0.00,0.13,0.44}{##1}}}
\expandafter\def\csname PY@tok@nn\endcsname{\let\PY@bf=\textbf\def\PY@tc##1{\textcolor[rgb]{0.05,0.52,0.71}{##1}}}
\expandafter\def\csname PY@tok@no\endcsname{\def\PY@tc##1{\textcolor[rgb]{0.38,0.68,0.84}{##1}}}
\expandafter\def\csname PY@tok@na\endcsname{\def\PY@tc##1{\textcolor[rgb]{0.25,0.44,0.63}{##1}}}
\expandafter\def\csname PY@tok@nb\endcsname{\def\PY@tc##1{\textcolor[rgb]{0.00,0.44,0.13}{##1}}}
\expandafter\def\csname PY@tok@nc\endcsname{\let\PY@bf=\textbf\def\PY@tc##1{\textcolor[rgb]{0.05,0.52,0.71}{##1}}}
\expandafter\def\csname PY@tok@nd\endcsname{\let\PY@bf=\textbf\def\PY@tc##1{\textcolor[rgb]{0.33,0.33,0.33}{##1}}}
\expandafter\def\csname PY@tok@ne\endcsname{\def\PY@tc##1{\textcolor[rgb]{0.00,0.44,0.13}{##1}}}
\expandafter\def\csname PY@tok@nf\endcsname{\def\PY@tc##1{\textcolor[rgb]{0.02,0.16,0.49}{##1}}}
\expandafter\def\csname PY@tok@si\endcsname{\let\PY@it=\textit\def\PY@tc##1{\textcolor[rgb]{0.44,0.63,0.82}{##1}}}
\expandafter\def\csname PY@tok@s2\endcsname{\def\PY@tc##1{\textcolor[rgb]{0.25,0.44,0.63}{##1}}}
\expandafter\def\csname PY@tok@vi\endcsname{\def\PY@tc##1{\textcolor[rgb]{0.73,0.38,0.84}{##1}}}
\expandafter\def\csname PY@tok@nt\endcsname{\let\PY@bf=\textbf\def\PY@tc##1{\textcolor[rgb]{0.02,0.16,0.45}{##1}}}
\expandafter\def\csname PY@tok@nv\endcsname{\def\PY@tc##1{\textcolor[rgb]{0.73,0.38,0.84}{##1}}}
\expandafter\def\csname PY@tok@s1\endcsname{\def\PY@tc##1{\textcolor[rgb]{0.25,0.44,0.63}{##1}}}
\expandafter\def\csname PY@tok@gp\endcsname{\let\PY@bf=\textbf\def\PY@tc##1{\textcolor[rgb]{0.78,0.36,0.04}{##1}}}
\expandafter\def\csname PY@tok@sh\endcsname{\def\PY@tc##1{\textcolor[rgb]{0.25,0.44,0.63}{##1}}}
\expandafter\def\csname PY@tok@ow\endcsname{\let\PY@bf=\textbf\def\PY@tc##1{\textcolor[rgb]{0.00,0.44,0.13}{##1}}}
\expandafter\def\csname PY@tok@sx\endcsname{\def\PY@tc##1{\textcolor[rgb]{0.78,0.36,0.04}{##1}}}
\expandafter\def\csname PY@tok@bp\endcsname{\def\PY@tc##1{\textcolor[rgb]{0.00,0.44,0.13}{##1}}}
\expandafter\def\csname PY@tok@c1\endcsname{\let\PY@it=\textit\def\PY@tc##1{\textcolor[rgb]{0.25,0.50,0.56}{##1}}}
\expandafter\def\csname PY@tok@kc\endcsname{\let\PY@bf=\textbf\def\PY@tc##1{\textcolor[rgb]{0.00,0.44,0.13}{##1}}}
\expandafter\def\csname PY@tok@c\endcsname{\let\PY@it=\textit\def\PY@tc##1{\textcolor[rgb]{0.25,0.50,0.56}{##1}}}
\expandafter\def\csname PY@tok@mf\endcsname{\def\PY@tc##1{\textcolor[rgb]{0.13,0.50,0.31}{##1}}}
\expandafter\def\csname PY@tok@err\endcsname{\def\PY@bc##1{\setlength{\fboxsep}{0pt}\fcolorbox[rgb]{1.00,0.00,0.00}{1,1,1}{\strut ##1}}}
\expandafter\def\csname PY@tok@kd\endcsname{\let\PY@bf=\textbf\def\PY@tc##1{\textcolor[rgb]{0.00,0.44,0.13}{##1}}}
\expandafter\def\csname PY@tok@ss\endcsname{\def\PY@tc##1{\textcolor[rgb]{0.32,0.47,0.09}{##1}}}
\expandafter\def\csname PY@tok@sr\endcsname{\def\PY@tc##1{\textcolor[rgb]{0.14,0.33,0.53}{##1}}}
\expandafter\def\csname PY@tok@mo\endcsname{\def\PY@tc##1{\textcolor[rgb]{0.13,0.50,0.31}{##1}}}
\expandafter\def\csname PY@tok@mi\endcsname{\def\PY@tc##1{\textcolor[rgb]{0.13,0.50,0.31}{##1}}}
\expandafter\def\csname PY@tok@kn\endcsname{\let\PY@bf=\textbf\def\PY@tc##1{\textcolor[rgb]{0.00,0.44,0.13}{##1}}}
\expandafter\def\csname PY@tok@o\endcsname{\def\PY@tc##1{\textcolor[rgb]{0.40,0.40,0.40}{##1}}}
\expandafter\def\csname PY@tok@kr\endcsname{\let\PY@bf=\textbf\def\PY@tc##1{\textcolor[rgb]{0.00,0.44,0.13}{##1}}}
\expandafter\def\csname PY@tok@s\endcsname{\def\PY@tc##1{\textcolor[rgb]{0.25,0.44,0.63}{##1}}}
\expandafter\def\csname PY@tok@kp\endcsname{\def\PY@tc##1{\textcolor[rgb]{0.00,0.44,0.13}{##1}}}
\expandafter\def\csname PY@tok@w\endcsname{\def\PY@tc##1{\textcolor[rgb]{0.73,0.73,0.73}{##1}}}
\expandafter\def\csname PY@tok@kt\endcsname{\def\PY@tc##1{\textcolor[rgb]{0.56,0.13,0.00}{##1}}}
\expandafter\def\csname PY@tok@sc\endcsname{\def\PY@tc##1{\textcolor[rgb]{0.25,0.44,0.63}{##1}}}
\expandafter\def\csname PY@tok@sb\endcsname{\def\PY@tc##1{\textcolor[rgb]{0.25,0.44,0.63}{##1}}}
\expandafter\def\csname PY@tok@k\endcsname{\let\PY@bf=\textbf\def\PY@tc##1{\textcolor[rgb]{0.00,0.44,0.13}{##1}}}
\expandafter\def\csname PY@tok@se\endcsname{\let\PY@bf=\textbf\def\PY@tc##1{\textcolor[rgb]{0.25,0.44,0.63}{##1}}}
\expandafter\def\csname PY@tok@sd\endcsname{\let\PY@it=\textit\def\PY@tc##1{\textcolor[rgb]{0.25,0.44,0.63}{##1}}}

\def\PYZus{\char`\_}
\def\PYZob{\char`\{}
\def\PYZcb{\char`\}}

\def\PYZam{\char`\&}
\def\PYZlt{\char`\<}
\def\PYZgt{\char`\>}
\def\PYZsh{\char`\#}

\def\PYZhy{\char`\-}
\def\PYZsq{\char`\'}
\def\PYZdq{\char`\"}

\makeatother

\providecommand*{\DUfootnotemark}[3]{%
  \raisebox{1em}{\hypertarget{#1}{}}%
  \hyperlink{#2}{\textsuperscript{#3}}%
}
\providecommand{\DUfootnotetext}[4]{%
  \begingroup%
  \renewcommand{\thefootnote}{%
    \protect\raisebox{1em}{\protect\hypertarget{#1}{}}%
    \protect\hyperlink{#2}{#3}}%
  \footnotetext{#4}%
  \endgroup%
}

\providecommand*{\DUrole}[2]{%
  \ifcsname DUrole#1\endcsname%
    \csname DUrole#1\endcsname{#2}%
  \else% backwards compatibility: try \docutilsrole#1{#2}
    \ifcsname docutilsrole#1\endcsname%
      \csname docutilsrole#1\endcsname{#2}%
    \else%
      #2%
    \fi%
  \fi%
}

\providecommand*{\DUroletitlereference}[1]{\textsl{#1}}

\ifthenelse{\isundefined{\hypersetup}}{
  \usepackage[colorlinks=true,linkcolor=blue,urlcolor=blue]{hyperref}
  \urlstyle{same} % normal text font (alternatives: tt, rm, sf)
}{}

\begin{document}
\newcounter{footnotecounter}\title{Enhancing SfePy with Isogeometric Analysis}\author{Robert Cimrman$^{\setcounter{footnotecounter}{1}\fnsymbol{footnotecounter}\setcounter{footnotecounter}{2}\fnsymbol{footnotecounter}}$%
          \setcounter{footnotecounter}{1}\thanks{\fnsymbol{footnotecounter} %
          Corresponding author: \protect\href{mailto:cimrman3@ntc.zcu.cz}{cimrman3@ntc.zcu.cz}}\setcounter{footnotecounter}{2}\thanks{\fnsymbol{footnotecounter} New Technologies Research Centre, University of West Bohemia,
Plzeň, Czech Republic}\thanks{%

          \noindent%
          Copyright\,\copyright\,2014 Robert Cimrman. This is an open-access article distributed under the terms of the Creative Commons Attribution License, which permits unrestricted use, distribution, and reproduction in any medium, provided the original author and source are credited. http://creativecommons.org/licenses/by/3.0/%
        }}\maketitle
          \renewcommand{\leftmark}{PROC. OF THE 7th EUR. CONF. ON PYTHON IN SCIENCE (EUROSCIPY 2014)}
          \renewcommand{\rightmark}{ENHANCING SFEPY WITH ISOGEOMETRIC ANALYSIS}

\setcounter{page}{65}
\newcommand*{\docutilsroleref}{\ref}
\newcommand*{\docutilsrolelabel}{\label}
\AtEndDocument{\cleardoublepage}
\begin{abstract}In the paper a recent enhancement to the open source package SfePy (Simple
Finite Elements in Python, \url{http://sfepy.org}) is introduced, namely the
addition of another numerical discretization scheme, the isogeometric
analysis, to the original implementation based on the nowadays standard and
well-established numerical solution technique, the finite element method.
The isogeometric removes the need of the solution domain approximation by a
piece-wise polygonal domain covered by the finite element mesh, and allows
approximation of unknown fields with a higher smoothness then the finite
element method, which can be advantageous in many applications. Basic
numerical examples illustrating the implementation and use of the
isogeometric analysis in SfePy are shown.\end{abstract}\begin{IEEEkeywords}partial differential equations, finite element method, isogeometric
analysis, SfePy\end{IEEEkeywords}

\section{Introduction%
  \label{introduction}%
}

Many problems in physics, biology, chemistry, geology and other scientific
disciplines can be described mathematically using a partial differential
equation (PDE) or a system of several PDEs. The PDEs are formulated in terms of
unknown field variables or fields, defined in some domain with a sufficiently
smooth boundary embedded in physical space.

SfePy (Simple Finite Elements in Python, \url{http://sfepy.org}) is a framework for
solving various kinds of problems (mechanics, physics, biology, ...) described
by PDEs in two or three space dimensions. Because only the most basic PDEs on
simple domains (circle, square, etc.) can be solved analytically, a numerical
solution scheme is needed, involving, typically:%
\begin{itemize}

\item 

an approximation of the original domain by a polygonal domain;
\item 

an approximation of continuous fields by discrete fields defined by a finite
set of degrees of freedom (DOFs) and a (piece-wise) polynomial basis.
\end{itemize}

The above steps are called \emph{discretization} of the continuous problem. In the
following text two discretization schemes will be briefly outlined:%
\begin{itemize}

\item 

the finite element method \cite{FEM} - a long-established industry
approved method based on piece-wise polynomial approximation,
\item 

the isogeometric analysis \cite{IGA} - a quite recent generalization of FEM that
uses spline- or NURBS-based approximation.
\end{itemize}

SfePy, as its name suggests, has been based on FEM from its very beginning. The
IGA implementation has been added mainly due to the following reasons (both
will be addressed more in the text):%
\begin{itemize}

\item 

The IGA approximation can be globally smooth on a single patch geometry. The
continuity is determined by a few well defined parameters. This fact was the
main factor in deciding to implement IGA, because the smoothness is crucial
in one of our research applications (ab-initio electronic structure
calculations - work in progress). The high smoothness is paid for by the
higher computational complexity of the NURBS basis evaluation and higher
fill-in of the sparse matrix that a problem discretization leads to.
\item 

IGA can work directly with the geometric description of objects used in
geometric modeling and computer-aided design (CAD) systems, removing thus the
meshing step.
\end{itemize}

The paper is structured as follows. The geometric representation of objects is
outlined in \hyperref[geometry-description-using-nurbs]{Geometry Description using NURBS}, because the terms defined
there are used in the IGA part of \hyperref[outline-of-fem-and-iga]{Outline of FEM and IGA}. Then the
particular choices made in SfePy are presented in \hyperref[iga-implementation-in-sfepy]{IGA Implementation in
SfePy} and illustrated using examples of PDE solutions in \hyperref[examples]{Examples}. All
computations below were done in SfePy, version 2014.3 - note that this paper is
a short description of the state and capabilities of the code as of this
version. The examples of numerical solutions have no particular scientific
meaning or importance besides being an illustration of the used methods.

\section{Geometry Description using NURBS%
  \label{geometry-description-using-nurbs}%
}

First, let us briefly review the geometric representation of objects using
Bézier curves, B-splines and \cite{NURBS} (Non-uniform rational B-spline) curves
and 2D (surface) or 3D (solid) bodies, to elucidate terminology used in
subsequent sections.  Our IGA implementation is based on the explanation and
algorithms in \cite{BE}, thus below we follow its notation and definitions.

\subsection{Bézier Curves%
  \label{bezier-curves}%
}

A Bézier curve is a parametric curve frequently used in computer graphics and
related fields. We define it here because its polynomial basis is used in the
code by means of the Bézier extraction technique, see \cite{BE} and below.

A degree $p$ Bézier curve is defined by a linear combination of
$p + 1$ \emph{Bernstein polynomial basis} functions as\begin{equation*}
C(\xi) = \sum_{a=1}^{p+1} \bm{P}_a B_{a,p}(\xi) = \bm{P}^T \bm{B}(\xi)
\mbox{ for } \xi \in [0, 1] \;,
\end{equation*}where $\bm{P} = \{\bm{P}_a\}_{a=1}^{p+1}$ is the set of control points and
$\bm{B}(\xi) = \{B_{a,p}(\xi)\}_{a=1}^{p+1}$ is the set of Bernstein
polynomial basis functions. The Bernstein basis can be defined recursively for
$\xi \in [0, 1]$ as $B_{a,p}(\xi) = (1 - \xi) B_{a,p-1}(\xi) + \xi
B_{a-1,p-1}(\xi)$, $B_{1,0}(\xi) \equiv 1$, $B_{a,p}(\xi) \equiv 0$
if $a < 1$ or $a > p + 1$. An example of Bézier curve is shown in
Fig. \DUrole{ref}{bezier-curve}.\begin{figure}[bht]\noindent\makebox[\columnwidth][c]{\includegraphics[scale=0.40]{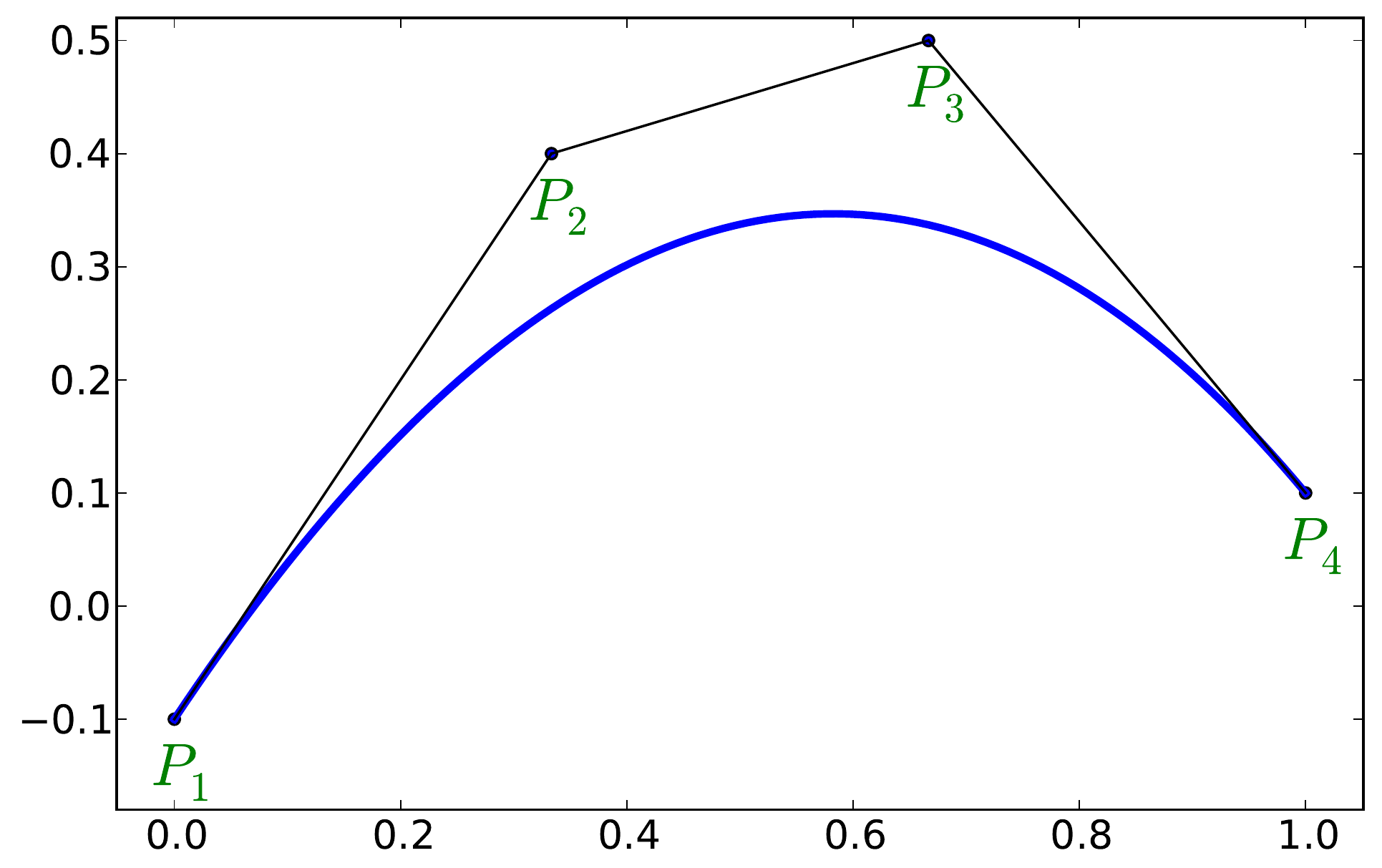}}
\caption{A Bézier curve (blue) of degree three with four control points in two space
dimensions. \DUrole{label}{bezier-curve}}
\end{figure}

\subsection{B-spline Curves%
  \label{b-spline-curves}%
}

A B-spline is a generalization of the Bézier curve. B-splines (and their NURBS
generalization, see below) are used in computer graphics, geometry modeling and
related fields as well as the Bézier curves. In IGA B-spline basis functions
can be used for approximation of the unknown fields.

A univariate B-spline curve of degree $p$ is defined by a linear
combination of $n$ basis functions as\begin{equation*}
T(\xi) = \sum_{A=1}^{n} \bm{P}_A N_{A,p}(\xi) = \bm{P}^T \bm{N}(\xi) \;,
\end{equation*}where $\bm{P} = \{\bm{P}_A\}_{A=1}^{n}$ is the set of control points. The
basis functions are defined by a \emph{knot vector} - a set of non-decreasing
parametric coordinates $\Xi = \{\xi_1, \xi_2, \dots, \xi_{n + p + 1}\}$,
where $\xi_A \in \mathbb{R}$ is the $A^{th}$ knot and $p$ is
the polynomial degree of the B-spline basis functions. Then for $p = 0$\begin{eqnarray*}
N_{A,0}(\xi) &=& 1 \mbox { for } \xi_A \leq \xi < \xi_{A+1} \;, \\
&=& 0 \mbox{ otherwise.}
\end{eqnarray*}For $p > 0$ the basis functions are defined by the Cox-de Boor recursion
formula\begin{equation*}
N_{A,p}(\xi) = \frac{\xi - \xi_A}{\xi_{A+p} - \xi_A} N_{A,p-1}(\xi)
+ \frac{\xi_{A+p+1} - \xi}{\xi_{A+p+1} - \xi_{A+1}}N_{A+1,p-1}(\xi) \;.
\end{equation*}Note that it is possible to insert knots into a knot vector without changing
the geometric or parametric properties of the curve by computing the new set of
control points in a particular way, see e.g. \cite{BE}.

A B-spline curve with a knot vector with no internal knots, i.e. of the form\begin{equation*}
\Xi = \{\underbrace{0, \dots, 0}_{p+1}, \underbrace{1, \dots, 1}_{p+1}\} \;,
\end{equation*}corresponds to a Bézier curve of degree $p$ with the same control points.

\subsection{NURBS Curves%
  \label{nurbs-curves}%
}

B-splines can be used to approximately describe almost any geometry. Their main
drawback is the fact, that a circular or spherical segment cannot be described
exactly. This problem was eliminated by the introduction of NURBS in geometry
modelling.

A NURBS (Non-uniform rational B-spline) of degree $p$ is defined by a
linear combination of $n$ rational basis functions as\begin{equation*}
T(\xi) = \sum_{A=1}^{n} \bm{P}_A R_{A,p}(\xi) = \bm{P}^T \bm{R}(\xi) \;,
\end{equation*}where $\bm{P} = \{\bm{P}_A\}_{A=1}^{n}$ is the set of control points and
$\bm{R}(\xi) = \{R_{A,p}(\xi)\}_{A=1}^{p+1}$ is the set of rational basis
functions. The rational basis functions are defined using the B-spline basis
functions as\begin{equation*}
R_{A,p}(\xi) = \frac{w_A {N_{A,p}(\xi)}}{W(\xi)} \;, \quad
W(\xi) = \sum_{B=1}^{n} w_B N_{B,p}(\xi) \;,
\end{equation*}where $w_i$ is the weight corresponding to the $i^{th}$ basis
function and $W$ is the weight function.

Note that a NURBS curve in $\mathbb{R}^n$ is equal to a B-spline curve in
$\mathbb{R}^{n+1}$:\begin{equation*}
T(\xi) = \sum_{A=1}^{n} \bar{\bm{P}_A} N_{A,p}(\xi) \;, \quad
\bar{\bm{P}_A} = \{w_A \bm{P}_A, w_A\}^T \;.
\end{equation*}This means that all algorithms that work for B-splines work also for NURBS.

\subsubsection{NURBS Surfaces and Solids%
  \label{nurbs-surfaces-and-solids}%
}

A surface is obtained by the tensor product of two NURBS curves.  The knot
vector is defined for each axial direction and there are $n \times m$
control points for $n$ basis functions in the first axis and $m$
basis functions in the second one.

Analogically, a solid is given by tensor product of three NURBS curves.

\subsubsection{NURBS Patches%
  \label{nurbs-patches}%
}

Complex geometries cannot be described by a single NURBS outlined above, often
called \emph{NURBS patch} - many such patches might be needed, and special care must
be taken to ensure required continuity along patch boundaries and to avoid
holes. A single patch geometry will be used in the following text, see
Fig. \DUrole{ref}{domain}.

\section{Outline of FEM and IGA%
  \label{outline-of-fem-and-iga}%
}

The two discretization methods will be illustrated on a very simple PDE - the
Laplace equation - in a plane (2D) domain. The Laplace equation describes
diffusion and can be used to determine, for example, temperature or electrical
potential distribution in the domain. We will use the \textquotedbl{}temperature\textquotedbl{}
terminology and the notation from Table \DUrole{ref}{notation}.\begin{table}
\setlength{\DUtablewidth}{0.8\linewidth}
\begin{longtable*}[c]{|p{0.274\DUtablewidth}|p{0.666\DUtablewidth}|}
\hline

symbol & 

meaning \\
\hline

$\Omega$ & 

solution domain \\
\hline

$\Omega_h$ & 

discretized solution domain \\
\hline

$\Gamma_D$, $\Gamma_N$ & 

subdomains representing parts of the
domain surface for applying Dirichlet and Neumann boundary conditions \\
\hline

$\underline{n}$ & 

unit outward normal \\
\hline

$\nabla \equiv [\frac{\partial}{\partial x_1},
\frac{\partial}{\partial x_2}]^T$ & 

gradient operator \\
\hline

$\nabla \cdot$ & 

divergence operator \\
\hline

$\Delta \equiv \nabla \cdot \nabla$ & 

Laplace operator \\
\hline

$C^1$ & 

space of functions with continuous first derivatives \\
\hline

$H^1$ & 

space of functions with integrable values and first derivatives \\
\hline

$H^1_0$ & 

space of functions from $H^1$ that are zero on
$\Gamma_D$ \\
\hline
\end{longtable*}
\caption{Notation. \DUrole{label}{notation}}\end{table}

The problem is as follows: Find temperature $T$ such that:\begin{eqnarray}
\label{strong}
\Delta T &=& 0 \mbox{ in } \Omega \;, \\
       T &=& \bar{T} \mbox{ on } \Gamma_D \;, \\
       \nabla T \cdot \underline{n} &=& 0 \mbox{ on } \Gamma_N \;,
\end{eqnarray}where the second equation is the Dirichlet (or essential) boundary condition
and the third equation is the Neumann (or natural) boundary condition that
corresponds to a flux through the boundary.

The operator $\Delta$ has second derivatives - that means that the
solution $T$ needs to have continuous first derivatives, or, it has to be
from $C^1$ function space - this is often not possible in examples from
practice. Instead, a \emph{weak solution} is sought that satisfies: Find $T
\in H^1(\Omega)$\begin{eqnarray}
\label{weak}
 \int_{\Omega} \nabla s \cdot \nabla T
 - \underbrace{\int_{\Gamma_N} s\ \nabla T \cdot \underline{n}}_{\equiv 0}
 &=& 0
 \;, \quad \forall s \in H^1_0(\Omega) \;, \\
 T &=& \bar{T} \quad \mbox{ on } \Gamma_D \;,
\end{eqnarray}where the natural boundary condition appears naturally in the equation (hence
its name). The above system can be derived by multiplying the original equation
by a test function $s \in H^1_0(\Omega)$, integrating over the whole
domain and then integrating by parts.

Both FEM and IGA now replace the infinite function space $H^1(\Omega)$ by
a finite subspace with a basis with a small support on a discretized domain
$\Omega_h$, see below for particular basis choices. Then
$T(\underline{x}) \approx \sum_{k=1}^{N} T_k \phi_k(\underline{x})$,
where $T_k$ are the DOFs and $\phi_k$ are the base
functions. Similarly, $s(\underline{x}) \approx \sum_{k=1}^{N} s_k
\phi_k(\underline{x})$. Substituting those into (\DUrole{ref}{weak}) we obtain\begin{eqnarray*}
\int_{\Omega_h} \left( \sum_{j=1}^{N} s_j \nabla \phi_j \cdot
\sum_{k=1}^{N} \nabla \phi_k T_k \right) = 0 \;.
\end{eqnarray*}This has to hold for any $s$, so we can choose $s = \phi_j$ for
$j = 1, \dots, N$. It is also possible to switch the sum with the
integral and put the constants $T_k$ out of the integral, to obtain the
discrete system:\begin{eqnarray}
\label{discrete}
\sum_{k=1}^{N} \int_{\Omega_h} \left(\nabla \phi_j \cdot
\nabla \phi_k \right) T_k = 0 \;.
\end{eqnarray}In compact matrix notation we can write $\bm{K} \bm{T} = \bm{0}$, where
the matrix $\bm{K}$ has components $K_{ij} = \int_{\Omega_h}
\nabla \phi_i \cdot \nabla \phi_j$ and $\bm{T}$ is the vector of
$T_k$. The Dirichlet boundary conditions are satisfied by setting the
$T_k$ on the boundary $\Gamma_D$ to appropriate values.

Both methods make use of the small support and evaluate (\DUrole{ref}{discrete}) as a
sum over small \textquotedbl{}elements\textquotedbl{} to obtain local matrices or vectors that are then
assembled into a global system - system of linear algebraic equations in our
case.

The particulars of domain geometry description and basis choice will now be
outlined. For both methods, we will use the domain shown in Figure
\DUrole{ref}{domain}. Its geometry is described by NURBS, see \hyperref[geometry-description-using-nurbs]{Geometry Description
using NURBS}.\begin{figure}[bht]\noindent\makebox[\columnwidth][c]{\includegraphics[scale=0.40]{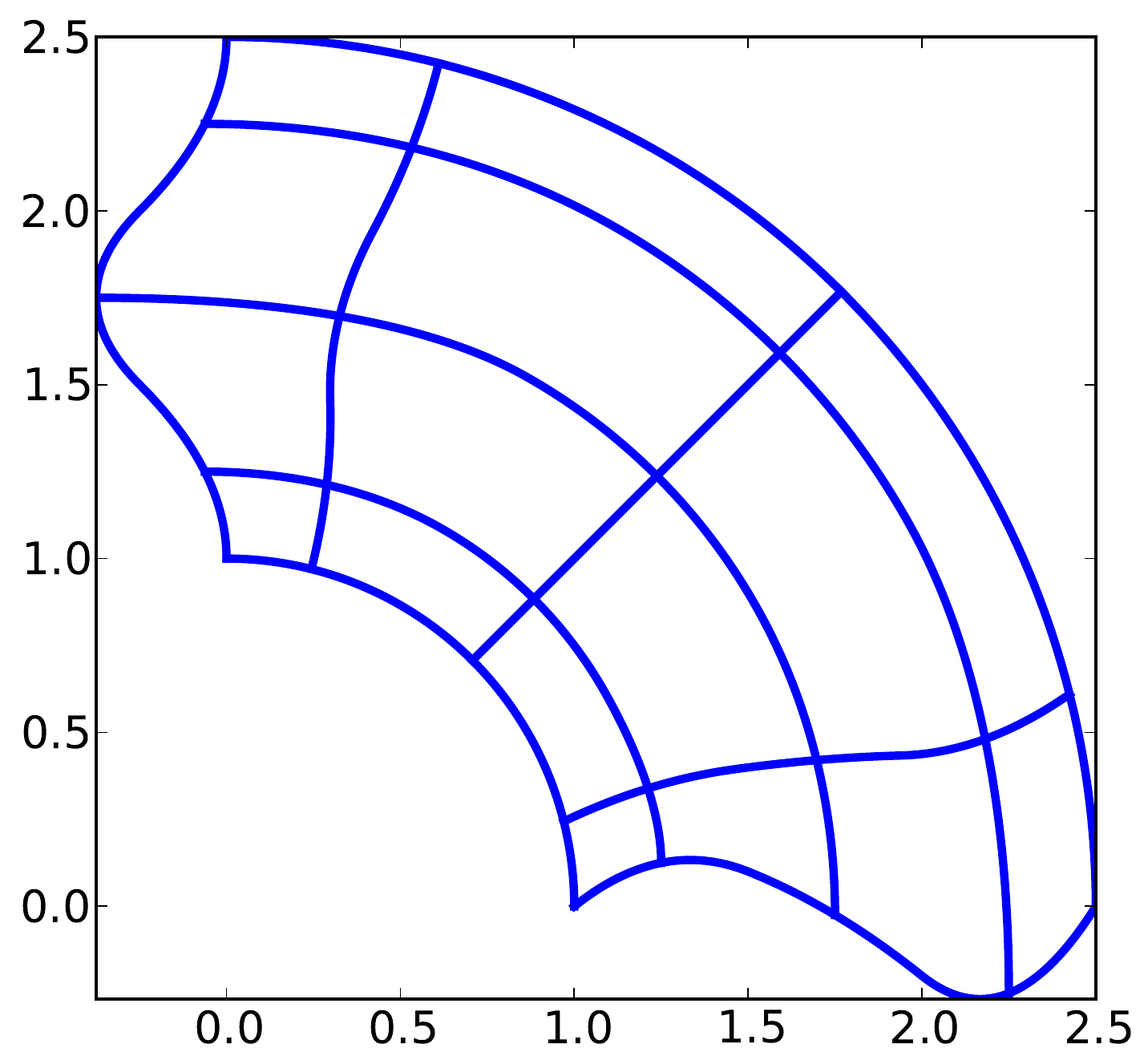}}
\caption{The domain with a curved boundary described by NURBS. The internal
lines correspond to several selected iso-lines given by the parametrization
of the 2D tensor-product NURBS patch. \DUrole{label}{domain}}
\end{figure}

\subsection{FEM%
  \label{fem}%
}

In this method a continuous solution domain is approximated by a polygonal
domain - \emph{FE mesh} - composed of small basic subdomains with a simple geometric
shape (e.g. triangles or quadrilaterals in 2D, tetrahedrons or hexahedrons in
3D) - the elements. The continuous fields of the PDEs are approximated by
polynomials defined on the individual elements. This approximation is (usually)
continuous over the whole domain, but its derivatives are only piece-wise
continuous.

First we need to make a FE mesh from the NURBS description, usual in CAD
systems. While it is easy for our domain, it is a difficult task in general,
especially in 3D space. Here a cheat has been used and the mesh depicted in
Figure \DUrole{ref}{fe-domain} was generated from the NURBS description using the IGA
techniques described below. Quite a fine mesh had to be used to capture the
curved boundaries.\begin{figure}[bht]\noindent\makebox[\columnwidth][c]{\includegraphics[scale=0.40]{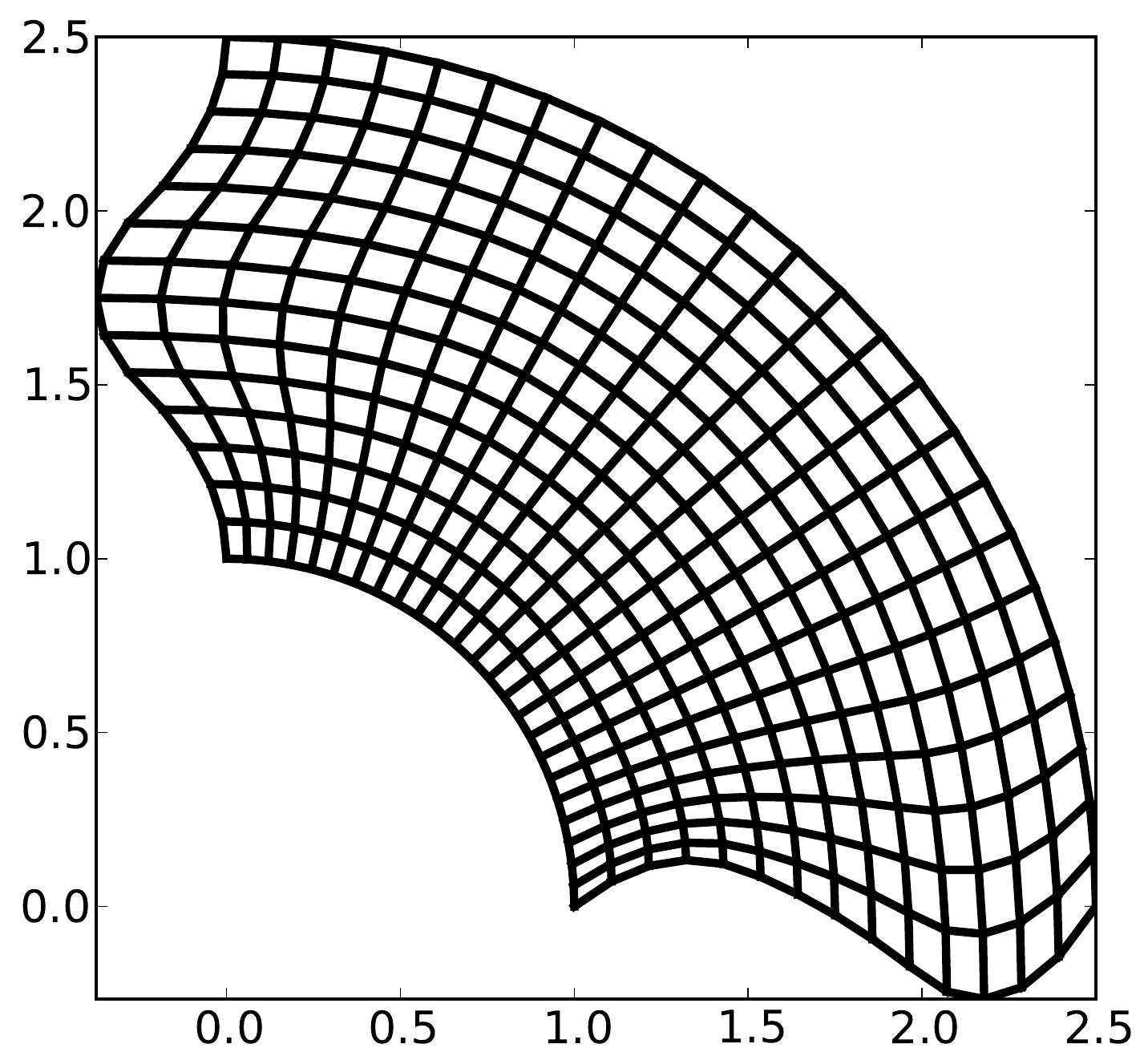}}
\caption{The FE-discretized domain covered by quadrilateral
elements, forming the FE mesh. \DUrole{label}{fe-domain}}
\end{figure}

Having the geometry discretized, a suitable approximation of the fields has to
be devised. In (classical\DUfootnotemark{id7}{id8}{1}) FEM, the base functions with small support are
polynomials, see Figure \DUrole{ref}{fe-basis-1d} for an illustration in 1D. A
$k$-th base function is nonzero only in elements that share the DOF
$T_k$ and it is a continuous polynomial over each element.%
\DUfootnotetext{id8}{id7}{1}{
See the Wikipedia page for a basic overview of FEM and its many
variations: \url{http://en.wikipedia.org/wiki/Finite_element_method}.}
\begin{figure}[bht]\noindent\makebox[\columnwidth][c]{\includegraphics[scale=0.30]{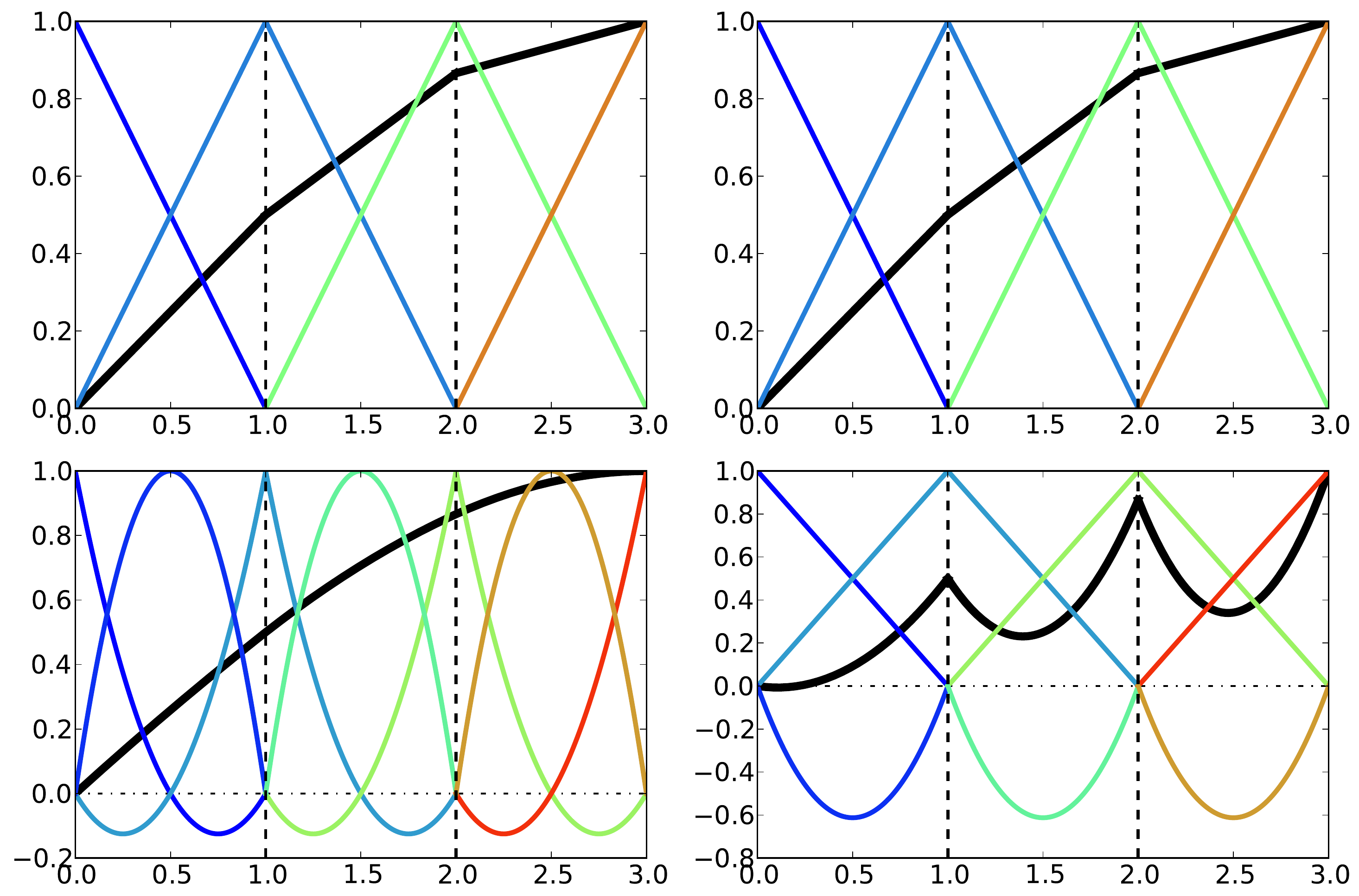}}
\caption{The 1D FE basis on three line elements with black thick line an interpolated
function resulting from the same DOF vector for each row: top: linear,
bottom: quadratic, left: Lagrange, right: Lobatto. Each basis function has a
single color. \DUrole{label}{fe-basis-1d}}
\end{figure}

The thick black lines in Figure \DUrole{ref}{fe-basis-1d} result from interpolation of
the DOF vector generated by $\sin(\frac{\pi}{2} \frac{x}{3})$ evaluated
in points of maximum of each basis function. The left column of the figure
shows the Lagrange polynomial basis, which is interpolatory, i.e., a DOF value
is equal to the approximated function value in the point, called \emph{node}, where
the basis is equal to 1. The right column of the figure shows the Lobatto
polynomial basis, that is not interpolatory for DOFs belonging to basis
functions with order greater than 1 - that is why the bottom right interpolated
function differs from the other cases. This complicates several things
(e.g. setting of Dirichlet boundary conditions - a projection is needed), but
the hierarchical nature of the basis, i.e. increasing approximation order means
adding new basis functions without modifying the existing ones, has also
advantages, for example better condition number of the matrix for higher order
approximations.

The basis functions are usually defined in a reference element, and are then
mapped to the physical mesh elements by an (affine) transformation. For our
mesh we will use bi-quadratic polynomials over the reference quadrilateral - a
quadratic function along each axis direction, such as those in the bottom row
of Figure \DUrole{ref}{fe-basis-1d}.

Several families of the element basis functions exist. In SfePy, Lagrange basis
and Lobatto (hierarchical) basis can be used on quadrilaterals, see Figure
\DUrole{ref}{fe-bases}.\begin{figure*}[]\noindent\makebox[\textwidth][c]{\includegraphics[scale=0.40]{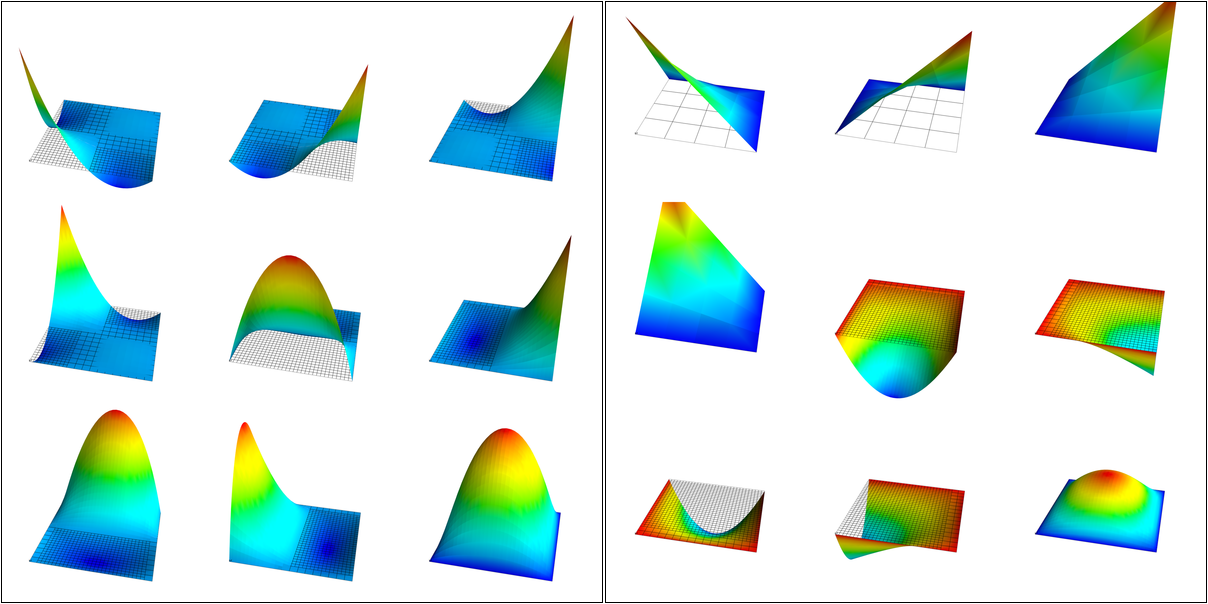}}
\caption{Bi-quadratic basis functions on the reference quadrilateral: left: Lagrange
right: Lobatto. \DUrole{label}{fe-bases}}
\end{figure*}

\subsection{IGA%
  \label{iga}%
}

In IGA, the CAD geometrical description in terms of NURBS patches is used
directly for the approximation of the unknown fields, without the intermediate
FE mesh - the meshing step is removed, which is one of its principal
advantages. As described in \hyperref[geometry-description-using-nurbs]{Geometry Description using NURBS}, a
D-dimensional geometric domain is defined by\begin{equation*}
\underline{x}(\underline{\xi})
= \sum_{A=1}^{n} \bm{P}_A R_{A,p}(\underline{\xi})
= \bm{P}^T \bm{R}(\underline{\xi}) \;,
\end{equation*}where $\underline{\xi} = \{\xi_1, \dots, \xi_D\}$ are the parametric
coordinates, and $\bm{P} = \{\bm{P}_A\}_{A=1}^{n}$ is the set of control
points. The same NURBS basis is used also for the approximation of PDE
solutions. For our temperature problem we have\begin{equation*}
T(\underline{\xi})
= \sum_{A=1}^{n} T_A R_{A,p}(\underline{\xi})\;,
\quad
s(\underline{\xi})
= \sum_{A=1}^{n} s_A R_{A,p}(\underline{\xi})\;,
\end{equation*}where $T_A$ are the unknown DOFs - coefficients of the basis in the linear
combination, and $s_A$ are the test function DOFs.\begin{figure*}[]\noindent\makebox[\textwidth][c]{\includegraphics[scale=0.50]{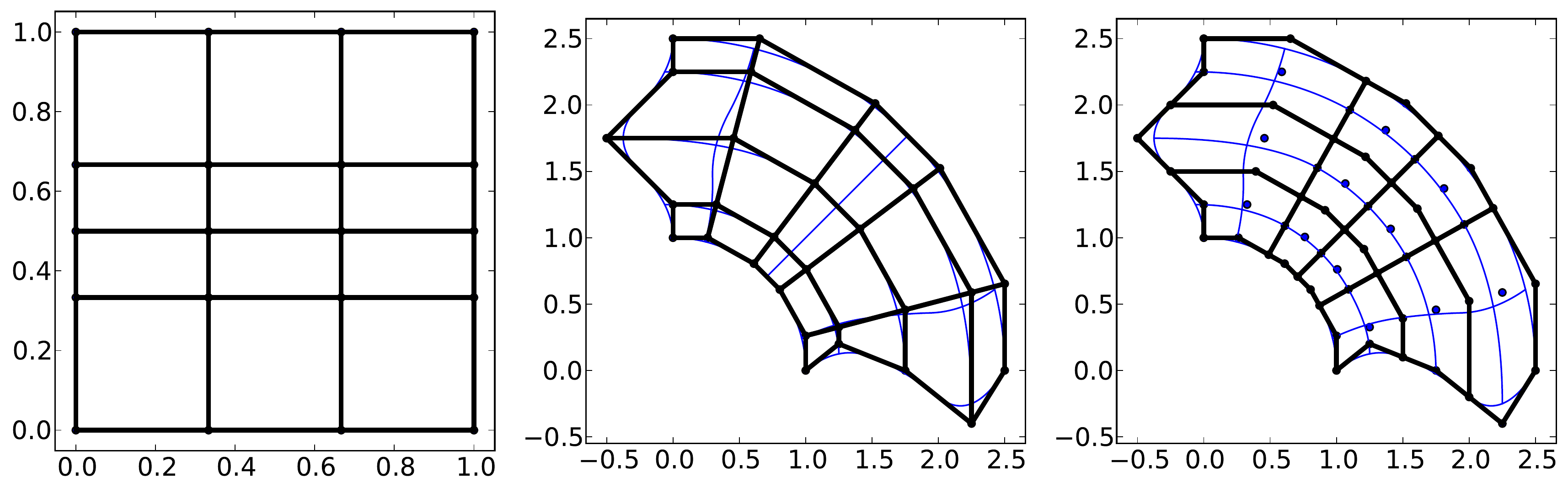}}
\caption{From left to right: parametric mesh (tensor product of knot vectors),
control mesh, Bézier mesh. \DUrole{label}{ig-domain-grids}}
\end{figure*}

Our domain in Figure \DUrole{ref}{domain} can be exactly described by a single NURBS
patch. Several auxiliary grids (called \textquotedbl{}meshes\textquotedbl{} as well, but do not mistake
with the FE mesh) can be drawn for the patch, see Figure
\DUrole{ref}{ig-domain-grids}. The parametric mesh is simply the tensor product of the
knot vectors defining the parametrization - the lines correspond to the knot
vector values. In our case there are four unique knot values in the first
parametric axis and five in the second axis. The control mesh has vertices
given by the NURBS patch control points and connectivity corresponding to the
tensor product nature of the patch. The Bézier mesh will be described below.
The thin blue lines are iso-lines of the NURBS parametrization, as in Figure
\DUrole{ref}{domain}.

On a single patch, such as our whole domain, the NURBS basis can be arbitrarily
smooth - this is another compelling feature not easily obtained by FEM. The
basis functions $R_{A,p}$, $A = 1, \dots, n$ on the patch are
uniquely determined by the knot vector for each axis, and cover the whole
patch, see Figure \DUrole{ref}{ig-base}.\begin{figure*}[]\noindent\makebox[\textwidth][c]{\includegraphics[scale=0.24]{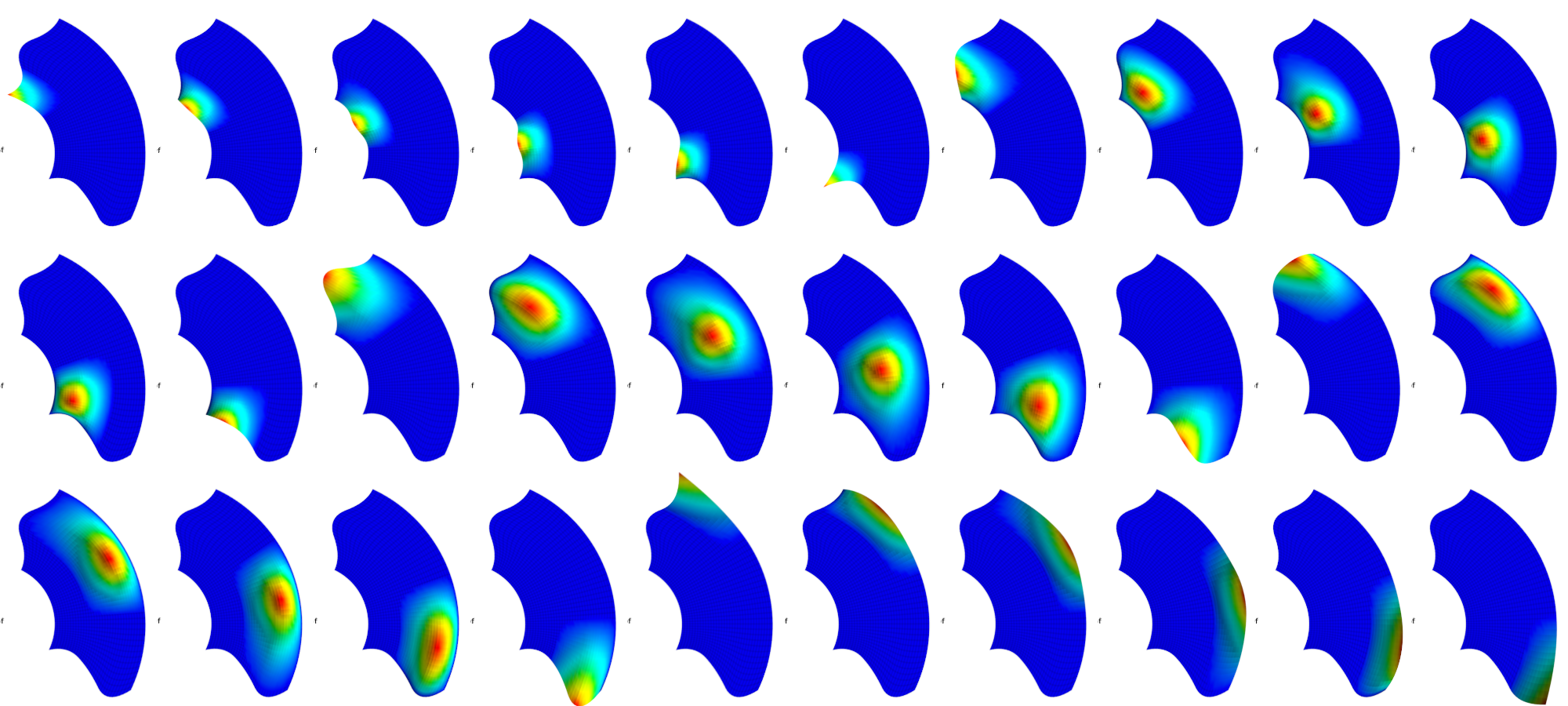}}
\caption{The degree 2 NURBS basis functions on the single patch
domain. \DUrole{label}{ig-base}}
\end{figure*}

\section{IGA Implementation in SfePy%
  \label{iga-implementation-in-sfepy}%
}

Our implementation uses a variant of IGA based on Bézier extraction operators
\cite{BE} that is suitable for inclusion into existing FE codes. The code itself
does not see the NURBS description at all. The NURBS description can be
prepared, for example, using \DUroletitlereference{igakit} package, a part of \cite{PetIGA}.

The Bézier extraction is illustrated in Figure \DUrole{ref}{bezier-extraction}. It is
based on the observation that repeating a knot in the knot vector decreases
continuity of the basis in that knot by one. This can be done in such a way
that the overall shape remains the same, but the \textquotedbl{}elements\textquotedbl{} appear naturally as
given by non-zero knot spans. The final basis restricted to each of the
elements is formed by the Bernstein polynomials $\bm{B}$.

In \cite{BE} algorithms are developed that allow computing \emph{Bézier extraction
operator} $\bm{C}$ for each such element such that the original (smooth)
NURBS basis function $\bm{R}$ can be recovered from the local Bernstein
basis $\bm{B}$ using $\bm{R} = \bm{C}\bm{B}$. The Bézier extraction
also allows construction of the Bézier mesh, see Figure \DUrole{ref}{ig-domain-grids},
right. The code then loops over the Bézier elements and assembles local
contributions in the usual FE sense.\begin{figure}[bht]\noindent\makebox[\columnwidth][c]{\includegraphics[scale=0.30]{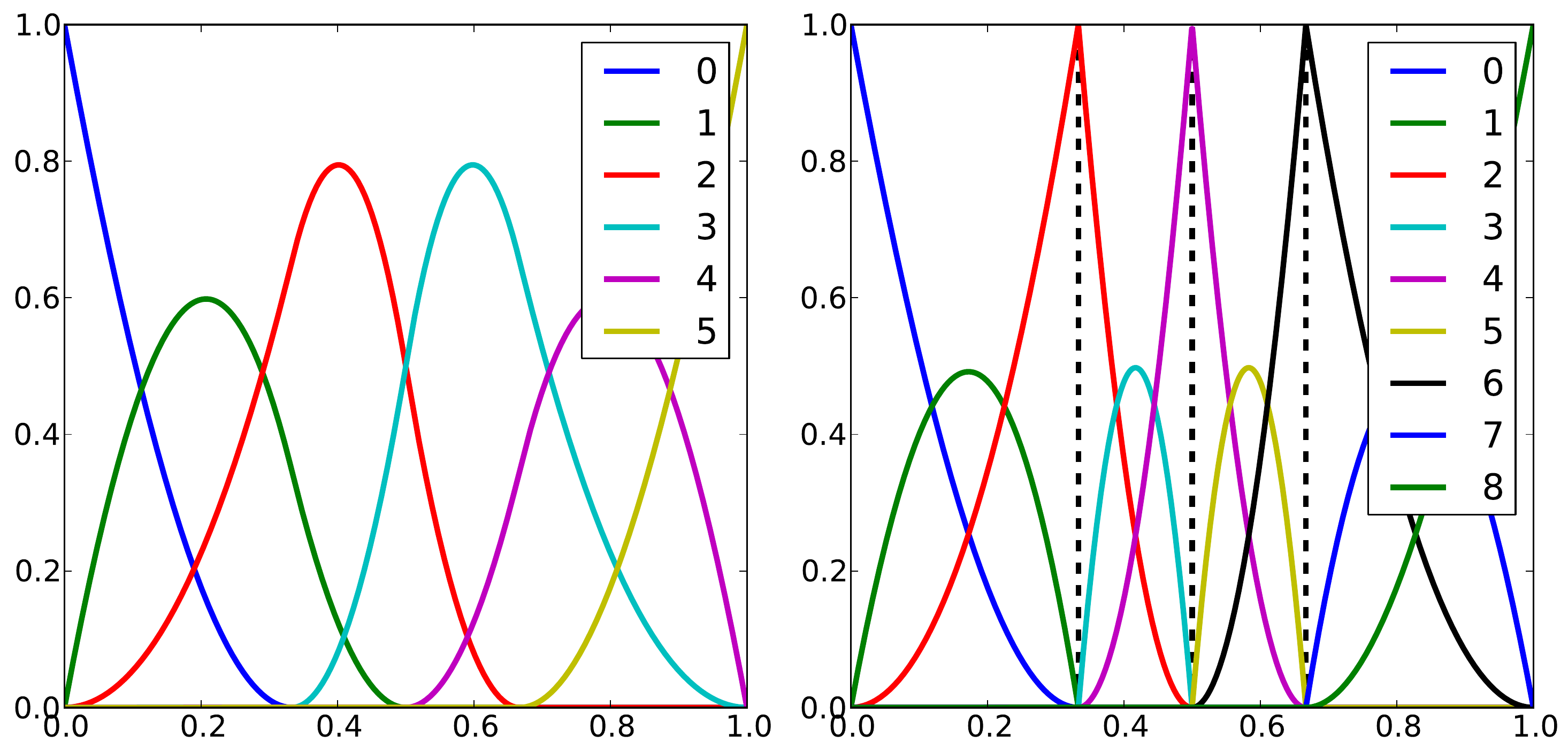}}
\caption{From left to right: NURBS basis of degree 2 that describes the second axis
of the parametric mesh, corresponding Bernstein basis with Bézier elements
delineated by vertical lines. \DUrole{label}{bezier-extraction}}
\end{figure}

In SfePy, various subdomains can be defined using \emph{regions}, see \cite{SfePy}. For
example, below we use the following region definition to specify an internal
subdomain:%
\begin{quote}\begin{verbatim}
'vertices in (x > 1.5) & (y < 1.5)'
\end{verbatim}

\end{quote}
To make this work with IGA, where no real mesh exists, a \emph{topological Bézier
mesh} is constructed, using only the corner vertices of the Bézier mesh
elements, because those are interpolatory, i.e., they are in the domain or on
its boundary, see Figures \DUrole{ref}{ig-domain-grids}, \DUrole{ref}{bezier-extraction}
right.

The regions serve both to specify integration domains of the terms that make up
the equations and to define the parts of boundary, where boundary conditions
are to be applied. SfePy supports setting the Dirichlet boundary conditions by
user-defined functions of space (and time). To make this feature work with IGA,
a projection of the boundary condition functions to the space spanned by the
appropriate boundary basis functions was implemented.

\subsection{Notes on Code Organization%
  \label{notes-on-code-organization}%
}

Although the Bézier extraction technique shields the IGA-specific code from the
rest of the FEM package, the implementation was not trivial. Similar to the
Lobatto FE basis, the DOFs corresponding to the NURBS basis are not equal to
function values with the exception of the patch corners. Moreover, the IGA
fields do not work with meshes at all - they need the NURBS description of the
domain together with the Bézier extraction operators and the topological Bézier
mesh. So the original \DUroletitlereference{sfepy.fem} sub-package was renamed and split into:%
\begin{itemize}

\item 

\DUroletitlereference{sfepy.discrete} for the general classes independent of the particular
discretization technique (for example variables, equations, boundary
conditions, materials, quadratures, etc.);
\item 

\DUroletitlereference{sfepy.discrete.fem} for the FEM-specific code;
\item 

\DUroletitlereference{sfepy.discrete.iga} for the IGA-specific code;
\item 

\DUroletitlereference{sfepy.discrete.common} for common functionality shared by some classes in
\DUroletitlereference{sfepy.discrete.fem} and \DUroletitlereference{sfepy.discrete.iga}.
\end{itemize}

In this way, circular import dependencies were minimized.

\subsection{Using IGA%
  \label{using-iga}%
}

As described in \cite{SfePy}, problems can be described either using problem
description files - Python modules containing definitions of the various
components (mesh, regions, fields, equations, ...)  using basic data types such
as \texttt{dict} and \texttt{tuple}, or using the \DUroletitlereference{sfepy} package classes directly
interactively or in a script. The former way is more basic and will be used in
the following.

In a FEM computation, a mesh has to be defined using:\begin{Verbatim}[commandchars=\\\{\},fontsize=\footnotesize]
\PY{n}{filename\PYZus{}mesh} \PY{o}{=} \PY{l+s}{\PYZsq{}}\PY{l+s}{fe\PYZus{}domain.mesh}\PY{l+s}{\PYZsq{}}
\end{Verbatim}
In an IGA computation, a NURBS domain has to be defined instead:\begin{Verbatim}[commandchars=\\\{\},fontsize=\footnotesize]
\PY{n}{filename\PYZus{}domain} \PY{o}{=} \PY{l+s}{\PYZsq{}}\PY{l+s}{ig\PYZus{}domain.iga}\PY{l+s}{\PYZsq{}}
\end{Verbatim}
where the \DUroletitlereference{'.iga'} suffix is used for a custom HDF5 file that can be prepared
by functions in \DUroletitlereference{sfepy.discrete.iga}.

A scalar real FE field with the approximation order 2 called 'temperature' can
be defined by:\begin{Verbatim}[commandchars=\\\{\},fontsize=\footnotesize]
\PY{c}{\PYZsh{} Lagrange basis is the default.}
\PY{n}{fields} \PY{o}{=} \PY{p}{\PYZob{}}
    \PY{l+s}{\PYZsq{}}\PY{l+s}{temperature}\PY{l+s}{\PYZsq{}} \PY{p}{:}
    \PY{p}{(}\PY{l+s}{\PYZsq{}}\PY{l+s}{real}\PY{l+s}{\PYZsq{}}\PY{p}{,} \PY{l+m+mi}{1}\PY{p}{,} \PY{l+s}{\PYZsq{}}\PY{l+s}{Omega}\PY{l+s}{\PYZsq{}}\PY{p}{,} \PY{l+m+mi}{2}\PY{p}{)}\PY{p}{,}
\PY{p}{\PYZcb{}}

\PY{c}{\PYZsh{} Lobatto basis.}
\PY{n}{fields} \PY{o}{=} \PY{p}{\PYZob{}}
    \PY{l+s}{\PYZsq{}}\PY{l+s}{temperature}\PY{l+s}{\PYZsq{}} \PY{p}{:}
    \PY{p}{(}\PY{l+s}{\PYZsq{}}\PY{l+s}{real}\PY{l+s}{\PYZsq{}}\PY{p}{,} \PY{l+m+mi}{1}\PY{p}{,} \PY{l+s}{\PYZsq{}}\PY{l+s}{Omega}\PY{l+s}{\PYZsq{}}\PY{p}{,} \PY{l+m+mi}{2}\PY{p}{,} \PY{l+s}{\PYZsq{}}\PY{l+s}{H1}\PY{l+s}{\PYZsq{}}\PY{p}{,} \PY{l+s}{\PYZsq{}}\PY{l+s}{lobatto}\PY{l+s}{\PYZsq{}}\PY{p}{)}\PY{p}{,}
\PY{p}{\PYZcb{}}
\end{Verbatim}
An analogical IGA field can be defined by:\begin{Verbatim}[commandchars=\\\{\},fontsize=\footnotesize]
\PY{n}{fields} \PY{o}{=} \PY{p}{\PYZob{}}
    \PY{l+s}{\PYZsq{}}\PY{l+s}{temperature}\PY{l+s}{\PYZsq{}} \PY{p}{:}
    \PY{p}{(}\PY{l+s}{\PYZsq{}}\PY{l+s}{real}\PY{l+s}{\PYZsq{}}\PY{p}{,} \PY{l+m+mi}{1}\PY{p}{,} \PY{l+s}{\PYZsq{}}\PY{l+s}{Omega}\PY{l+s}{\PYZsq{}}\PY{p}{,} \PY{n+nb+bp}{None}\PY{p}{,} \PY{l+s}{\PYZsq{}}\PY{l+s}{H1}\PY{l+s}{\PYZsq{}}\PY{p}{,} \PY{l+s}{\PYZsq{}}\PY{l+s}{iga}\PY{l+s}{\PYZsq{}}\PY{p}{)}\PY{p}{,}
\PY{p}{\PYZcb{}}
\end{Verbatim}
Here the approximation order is \DUroletitlereference{None}, as it is given by the \DUroletitlereference{'.iga'} domain
file.

The above are the only changes required to use IGA - everything else remains
the same as in FEM calculations. The scalar and vector volume terms (weak
forms, linear or nonlinear) listed at
\url{http://sfepy.org/doc-devel/terms_overview.html} can be used without
modification.

\subsection{Limitations%
  \label{limitations}%
}

There are currently several limitations that will be addressed in future:%
\begin{itemize}

\item 

projections of functions into the NURBS basis;
\item 

support for surface integrals;
\item 

linearization of results for post-processing;%
\begin{itemize}

\item 

currently the fields on a tensor-product patch are sampled by fixed
parameter vectors and a corresponding FE-mesh is generated;
\end{itemize}

\item 

all variables have to have the same approximation order, as the basis is
given by the domain file;
\item 

the domain is a single NURBS patch only.
\end{itemize}

\section{Examples%
  \label{examples}%
}

Numerical examples illustrating the IGA calculations are presented below. The
corresponding problem description files for SfePy (version 2014.3) can be
downloaded from \url{https://github.com/sfepy/euroscipy2014-iga-examples}, revision
28b1fe9bff7043da6fd159c20b9f244337a17e82, or from
\url{http://dx.doi.org/10.5281/zenodo.12257}.

\subsection{Temperature Distribution%
  \label{temperature-distribution}%
}

The 2D domain depicted in Figure \DUrole{ref}{domain} is used in this example.  The
temperature distribution is given by the solution of the Laplace equation
(\DUrole{ref}{weak}) with a particular set of Dirichlet boundary conditions on
$\Gamma_D$. The region $\Gamma_D$ consisted of two parts
$\Gamma_1$, $\Gamma_2$ of the domain boundary on the opposite edges
of the patch, see Figure \DUrole{ref}{domain-regions} - the temperature was fixed to
0.5 on $\Gamma_1$ and to -0.5 on $\Gamma_2$, as can be seen in
Figure \DUrole{ref}{laplace}. As mentioned in \hyperref[limitations]{Limitations}, the resulting field
$T$ was sampled by fixed uniform parameter vectors along each axis, and
the corresponding output FE mesh was generated. The mesh was saved in the VTK
format and the results visualized using SfePy's \texttt{postproc.py} script based on
Mayavi. The generated mesh can be seen as the undeformed wire-frame.\begin{figure}[bht]\noindent\makebox[\columnwidth][c]{\includegraphics[scale=0.30]{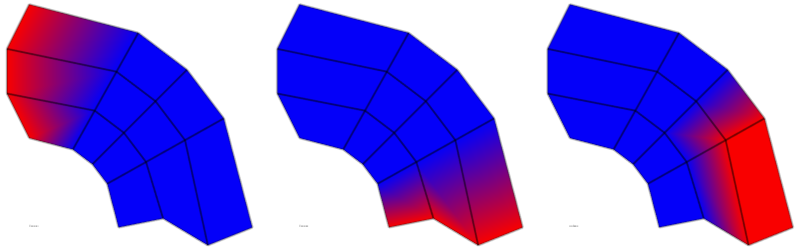}}
\caption{The regions defined on the domain shown on the topological Bézier mesh by
red color. From left: $\Gamma_1$, $\Gamma_2$, $\Omega_0$
\DUrole{label}{domain-regions}}
\end{figure}

For comparison with a FEM solution, see Figure \DUrole{ref}{laplace-fem}. The FEM
problem had 1363 DOFs in the linear system, while the IGA problem only 20. The
mesh depicted in Figure \DUrole{ref}{fe-domain} was used for the FEM computation.\begin{figure}[bht]\noindent\makebox[\columnwidth][c]{\includegraphics[scale=0.30]{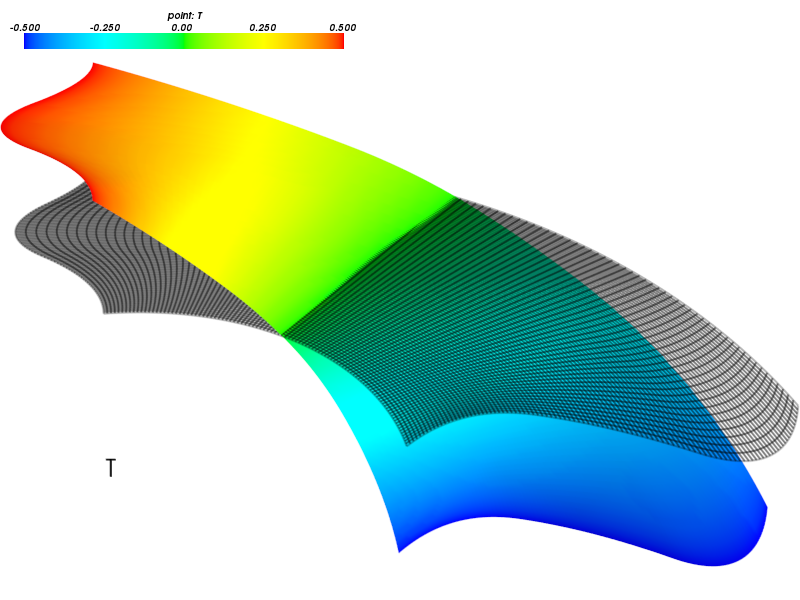}}
\caption{A solution of the 2D Laplace equation. \DUrole{label}{laplace}}
\end{figure}\begin{figure}[bht]\noindent\makebox[\columnwidth][c]{\includegraphics[scale=0.30]{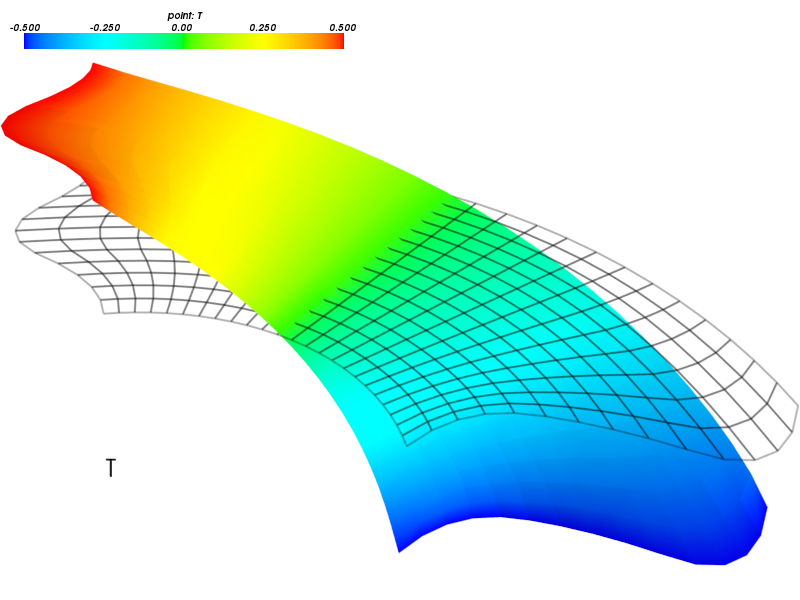}}
\caption{A solution of the 2D Laplace equation by FEM. \DUrole{label}{laplace-fem}}
\end{figure}

Next we added a negative source term to the Laplace equation in region
$\Omega_0$ (see Figure \DUrole{ref}{domain-regions} right) to obtain the Poisson
equation:\begin{eqnarray}
\label{weak-vf}
 \int_{\Omega} \nabla s \cdot \nabla T
 &=& \int_{\Omega_0} -2 s
 \;, \quad \forall s \in H^1_0(\Omega) \;, \\
 T &=& \bar{T} \quad \mbox{ on } \Gamma_D \;,
\end{eqnarray}The corresponding solution can be seen in Figure \DUrole{ref}{poisson}. The boundary
conditions stayed the same as in the previous case.\begin{figure}[bht]\noindent\makebox[\columnwidth][c]{\includegraphics[scale=0.30]{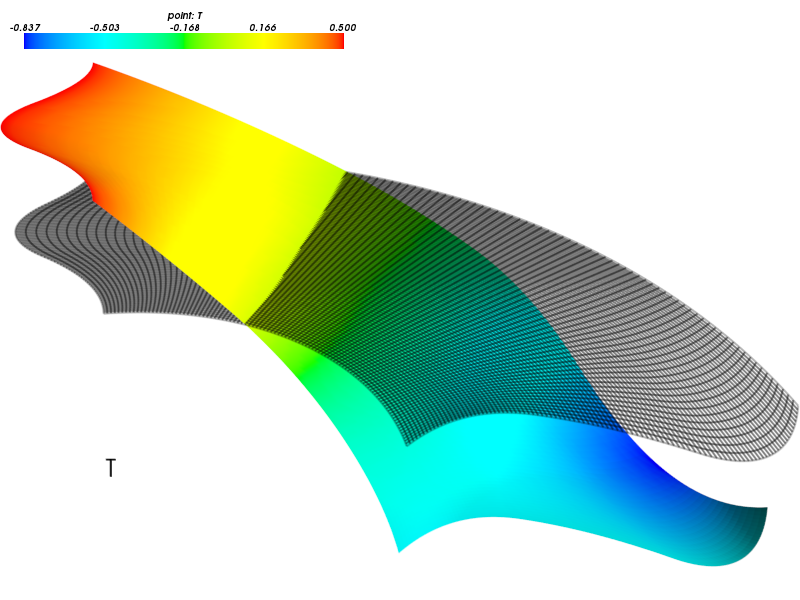}}
\caption{A solution of the 2D Poisson equation with volume source in a
subdomain. \DUrole{label}{poisson}}
\end{figure}

The complete problem description file for computing (\DUrole{ref}{weak-vf}) is shown
below. See \cite{SfePy} or \url{http://sfepy.org} for explanation.\begin{Verbatim}[commandchars=\\\{\},fontsize=\footnotesize]
\PY{n}{filename\PYZus{}domain} \PY{o}{=} \PY{l+s}{\PYZsq{}}\PY{l+s}{ig\PYZus{}domain.iga}\PY{l+s}{\PYZsq{}}

\PY{n}{regions} \PY{o}{=} \PY{p}{\PYZob{}}
    \PY{l+s}{\PYZsq{}}\PY{l+s}{Omega}\PY{l+s}{\PYZsq{}} \PY{p}{:} \PY{l+s}{\PYZsq{}}\PY{l+s}{all}\PY{l+s}{\PYZsq{}}\PY{p}{,}
    \PY{l+s}{\PYZsq{}}\PY{l+s}{Omega\PYZus{}0}\PY{l+s}{\PYZsq{}} \PY{p}{:} \PY{l+s}{\PYZsq{}}\PY{l+s}{vertices in (x \PYZgt{} 1.5) \PYZam{} (y \PYZlt{} 1.5)}\PY{l+s}{\PYZsq{}}\PY{p}{,}
    \PY{l+s}{\PYZsq{}}\PY{l+s}{Gamma1}\PY{l+s}{\PYZsq{}} \PY{p}{:} \PY{p}{(}\PY{l+s}{\PYZsq{}}\PY{l+s}{vertices of set xi10}\PY{l+s}{\PYZsq{}}\PY{p}{,} \PY{l+s}{\PYZsq{}}\PY{l+s}{facet}\PY{l+s}{\PYZsq{}}\PY{p}{)}\PY{p}{,}
    \PY{l+s}{\PYZsq{}}\PY{l+s}{Gamma2}\PY{l+s}{\PYZsq{}} \PY{p}{:} \PY{p}{(}\PY{l+s}{\PYZsq{}}\PY{l+s}{vertices of set xi11}\PY{l+s}{\PYZsq{}}\PY{p}{,} \PY{l+s}{\PYZsq{}}\PY{l+s}{facet}\PY{l+s}{\PYZsq{}}\PY{p}{)}\PY{p}{,}
\PY{p}{\PYZcb{}}

\PY{n}{fields} \PY{o}{=} \PY{p}{\PYZob{}}
    \PY{l+s}{\PYZsq{}}\PY{l+s}{temperature}\PY{l+s}{\PYZsq{}}
    \PY{p}{:} \PY{p}{(}\PY{l+s}{\PYZsq{}}\PY{l+s}{real}\PY{l+s}{\PYZsq{}}\PY{p}{,} \PY{l+m+mi}{1}\PY{p}{,} \PY{l+s}{\PYZsq{}}\PY{l+s}{Omega}\PY{l+s}{\PYZsq{}}\PY{p}{,} \PY{n+nb+bp}{None}\PY{p}{,} \PY{l+s}{\PYZsq{}}\PY{l+s}{H1}\PY{l+s}{\PYZsq{}}\PY{p}{,} \PY{l+s}{\PYZsq{}}\PY{l+s}{iga}\PY{l+s}{\PYZsq{}}\PY{p}{)}\PY{p}{,}
\PY{p}{\PYZcb{}}

\PY{n}{variables} \PY{o}{=} \PY{p}{\PYZob{}}
    \PY{l+s}{\PYZsq{}}\PY{l+s}{T}\PY{l+s}{\PYZsq{}} \PY{p}{:} \PY{p}{(}\PY{l+s}{\PYZsq{}}\PY{l+s}{unknown field}\PY{l+s}{\PYZsq{}}\PY{p}{,} \PY{l+s}{\PYZsq{}}\PY{l+s}{temperature}\PY{l+s}{\PYZsq{}}\PY{p}{,} \PY{l+m+mi}{0}\PY{p}{)}\PY{p}{,}
    \PY{l+s}{\PYZsq{}}\PY{l+s}{s}\PY{l+s}{\PYZsq{}} \PY{p}{:} \PY{p}{(}\PY{l+s}{\PYZsq{}}\PY{l+s}{test field}\PY{l+s}{\PYZsq{}}\PY{p}{,}    \PY{l+s}{\PYZsq{}}\PY{l+s}{temperature}\PY{l+s}{\PYZsq{}}\PY{p}{,} \PY{l+s}{\PYZsq{}}\PY{l+s}{T}\PY{l+s}{\PYZsq{}}\PY{p}{)}\PY{p}{,}
\PY{p}{\PYZcb{}}

\PY{n}{ebcs} \PY{o}{=} \PY{p}{\PYZob{}}
    \PY{l+s}{\PYZsq{}}\PY{l+s}{T1}\PY{l+s}{\PYZsq{}} \PY{p}{:} \PY{p}{(}\PY{l+s}{\PYZsq{}}\PY{l+s}{Gamma1}\PY{l+s}{\PYZsq{}}\PY{p}{,} \PY{p}{\PYZob{}}\PY{l+s}{\PYZsq{}}\PY{l+s}{T.0}\PY{l+s}{\PYZsq{}} \PY{p}{:} \PY{l+m+mf}{0.5}\PY{p}{\PYZcb{}}\PY{p}{)}\PY{p}{,}
    \PY{l+s}{\PYZsq{}}\PY{l+s}{T2}\PY{l+s}{\PYZsq{}} \PY{p}{:} \PY{p}{(}\PY{l+s}{\PYZsq{}}\PY{l+s}{Gamma2}\PY{l+s}{\PYZsq{}}\PY{p}{,} \PY{p}{\PYZob{}}\PY{l+s}{\PYZsq{}}\PY{l+s}{T.0}\PY{l+s}{\PYZsq{}} \PY{p}{:} \PY{o}{\PYZhy{}}\PY{l+m+mf}{0.5}\PY{p}{\PYZcb{}}\PY{p}{)}\PY{p}{,}
\PY{p}{\PYZcb{}}

\PY{n}{materials} \PY{o}{=} \PY{p}{\PYZob{}}
    \PY{l+s}{\PYZsq{}}\PY{l+s}{m}\PY{l+s}{\PYZsq{}} \PY{p}{:} \PY{p}{(}\PY{p}{\PYZob{}}\PY{l+s}{\PYZsq{}}\PY{l+s}{f}\PY{l+s}{\PYZsq{}} \PY{p}{:} \PY{o}{\PYZhy{}}\PY{l+m+mf}{2.0}\PY{p}{\PYZcb{}}\PY{p}{,}\PY{p}{)}\PY{p}{,}
\PY{p}{\PYZcb{}}

\PY{n}{integrals} \PY{o}{=} \PY{p}{\PYZob{}}
    \PY{l+s}{\PYZsq{}}\PY{l+s}{i}\PY{l+s}{\PYZsq{}} \PY{p}{:} \PY{l+m+mi}{3}\PY{p}{,}
\PY{p}{\PYZcb{}}

\PY{n}{equations} \PY{o}{=} \PY{p}{\PYZob{}}
    \PY{l+s}{\PYZsq{}}\PY{l+s}{Temperature}\PY{l+s}{\PYZsq{}}
    \PY{p}{:} \PY{l+s}{\PYZdq{}\PYZdq{}\PYZdq{}}\PY{l+s}{dw\PYZus{}laplace.i.Omega(s, T)}
\PY{l+s}{       = dw\PYZus{}volume\PYZus{}lvf.i.Omega\PYZus{}0(m.f, s)}\PY{l+s}{\PYZdq{}\PYZdq{}\PYZdq{}}
\PY{p}{\PYZcb{}}

\PY{n}{solvers} \PY{o}{=} \PY{p}{\PYZob{}}
    \PY{l+s}{\PYZsq{}}\PY{l+s}{ls}\PY{l+s}{\PYZsq{}} \PY{p}{:} \PY{p}{(}\PY{l+s}{\PYZsq{}}\PY{l+s}{ls.scipy\PYZus{}direct}\PY{l+s}{\PYZsq{}}\PY{p}{,} \PY{p}{\PYZob{}}\PY{p}{\PYZcb{}}\PY{p}{)}\PY{p}{,}
    \PY{l+s}{\PYZsq{}}\PY{l+s}{newton}\PY{l+s}{\PYZsq{}} \PY{p}{:} \PY{p}{(}\PY{l+s}{\PYZsq{}}\PY{l+s}{nls.newton}\PY{l+s}{\PYZsq{}}\PY{p}{,} \PY{p}{\PYZob{}}
        \PY{l+s}{\PYZsq{}}\PY{l+s}{i\PYZus{}max}\PY{l+s}{\PYZsq{}}      \PY{p}{:} \PY{l+m+mi}{1}\PY{p}{,}
        \PY{l+s}{\PYZsq{}}\PY{l+s}{eps\PYZus{}a}\PY{l+s}{\PYZsq{}}      \PY{p}{:} \PY{l+m+mf}{1e\PYZhy{}10}\PY{p}{,}
    \PY{p}{\PYZcb{}}\PY{p}{)}\PY{p}{,}
\PY{p}{\PYZcb{}}
\end{Verbatim}

\subsection{Elastic Deformation%
  \label{elastic-deformation}%
}
This example illustrates a calculation with a vector variable, the displacement
field $\underline{u}$, given by deformation of a 3D elastic body. The
weak form of the problem is: Find $\underline{u} \in [H^1(\Omega)]^3$
such that:\begin{eqnarray*}
 \int_{\Omega} D_{ijkl}\ e_{ij}(\underline{v}) e_{kl}(\underline{u})
 &=& 0
 \;, \quad \forall \underline{v} \in [H^1_0(\Omega)]^3 \;, \\
 \underline{u} &=& \bar{\underline{u}} \quad \mbox{ on } \Gamma_D \;,
\end{eqnarray*}where $D_{ijkl} = \mu (\delta_{ik} \delta_{jl}+\delta_{il} \delta_{jk}) +
\lambda \delta_{ij} \delta_{kl}$ is the isotropic stiffness tensor given in
terms of Lamé's coefficients $\lambda$, $\mu$ and
$e_{ij}(\underline{u}) = \frac{1}{2}(\frac{\partial u_i}{\partial x_j} +
\frac{\partial u_j}{\partial x_i})$ is the Cauchy, or small strain, deformation
tensor. The equation expresses the internal and external (zero here) force
balance, where the internal forces are described by the Cauchy stress tensor
$\sigma_{ij}(\underline{u}) = D_{ijkl}\ e_{kl}(\underline{u})$.

The 3D domain $\Omega$ was simply obtained by extrusion of the 2D domain
of the previous example, and again $\Gamma_D$ consisted of two parts
$\Gamma_1$, $\Gamma_2$. The body was clamped on $\Gamma_1$:
$\underline{u} = 0$ and displaced on $\Gamma_2$: $u_1 =
0.01$, $u_2(\underline{x}) = -0.02 x_2$ and $u_3(\underline{x}) =
-0.02 + (0.15 * (x_1 - 1)^2)$, for $\underline{x} \in \Gamma_2$. Note
that the Dirichlet boundary conditions on $\Gamma_2$ depend on the
position $\underline{x}$. The corresponding solution can be seen in
Figure \DUrole{ref}{elasticity}.\begin{figure}[bht]\noindent\makebox[\columnwidth][c]{\includegraphics[scale=0.30]{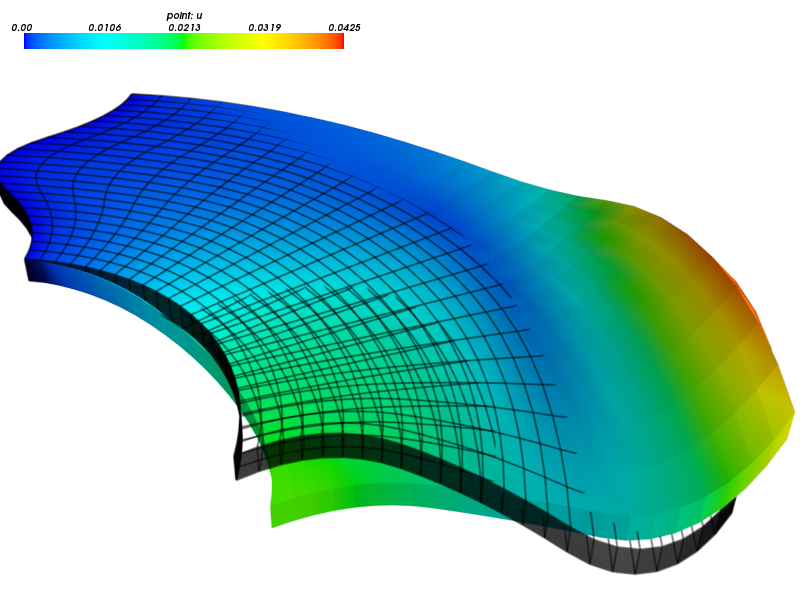}}
\caption{A solution of the 3D linear elasticity equation. The undeformed domain is
shown as a wireframe, 10x magnified deformation. \DUrole{label}{elasticity}}
\end{figure}

\section{Conclusion%
  \label{conclusion}%
}

Two numerical techniques for discretization of partial differential equations
were briefly outlined and compared, namely the well-established and proven
finite element method and its much more recent generalization, the isogeometric
analysis, on the background given by the open source finite element package
SfePy, that has been recently enhanced with the isogeometric analysis
functionality.

The Bézier extraction operators technique, that was used for a relatively
seamless integration into the existing finite element package, was mentioned,
as well as some of the difficulties \textquotedbl{}on the road\textquotedbl{} and limitations of the
current version.

Numerical examples - a scalar diffusion problem in 2D and a vector elastic body
deformation problem in 2D were shown.

\subsection{Support%
  \label{support}%
}

Work on SfePy is partially supported by the Grant Agency of the Czech Republic,
project P108/11/0853.
% Customised LaTeX packages

% -------------------------

% Please avoid using this feature, unless agreed upon with the

% proceedings editors.

% ::

% .. latex::

% :usepackage: somepackage

% Some custom LaTeX source here.

\end{document}